\documentclass[conference]{IEEEtran}
\IEEEoverridecommandlockouts

\usepackage{fancyhdr}

\usepackage{epstopdf}
\usepackage{cite}
\usepackage{amsmath,amssymb,amsfonts}
\usepackage[linesnumbered,ruled,vlined]{algorithm2e}
\usepackage{comment}
\usepackage{graphicx}
\usepackage{hyperref}
\usepackage{textcomp}

\usepackage{xcolor}
\def\BibTeX{{\rm B\kern-.05em{\sc i\kern-.025em b}\kern-.08em
    T\kern-.1667em\lower.7ex\hbox{E}\kern-.125emX}}

\newcommand{\kento}[1]{{\color{violet}[Kento:~#1]}}
\usepackage{url}

\newcommand{\az}[1]{{\color{brown}[az:~#1]}}

\begin{document}

\title{AutoCheck: Automatically Identifying Variables for Checkpointing by Data Dependency Analysis}


\author{%
  \IEEEauthorblockN{Xiang Fu$^1$, Weiping Zhang$^1$, Xin Huang$^1$, Wubiao Xu$^1$, Shiman Meng$^1$, Luanzheng Guo$^2$, Kento Sato$^3$}
  \IEEEauthorblockA{$^1$Nanchang Hangkong University, $^2$Pacific Northwest National Laboratory, $^3$RIKEN R-CCS\\
  \textit{\{fuxiang,wzhang,smeng,xhuang,wxu\}@nchu.edu.cn}; 
  \textit{lenny.guo@pnnl.gov}; 
  \textit{kento.sato@riken.jp};
  }
}

\maketitle

\thispagestyle{fancy}
\lhead{}
\rhead{}
\chead{}


\lfoot{\footnotesize{
SC24, November 17-22, 2024, Atlanta, Georgia, USA 
\newline 979-8-3503-5291-7/24/\$31.00 \copyright 2024 IEEE }}
\cfoot{}
\rfoot{}

\renewcommand{\headrulewidth}{0pt}
\renewcommand{\footrulewidth}{0pt}

\begin{abstract}
Checkpoint/Restart (C/R) has been widely deployed in numerous HPC systems, Clouds, and industrial data centers, which are typically operated by system engineers. Nevertheless, there is no existing approach that helps system engineers without domain expertise, and domain scientists without system fault tolerance knowledge identify those critical variables accounted for correct application execution restoration in a failure for C/R. To address this problem, we propose an analytical model and a tool (AutoCheck) that can automatically identify critical variables to checkpoint for C/R. AutoCheck relies on first, analytically tracking and optimizing data dependency between variables and other application execution state, and second, a set of heuristics that identify critical variables for checkpointing from the refined data dependency graph (DDG). AutoCheck allows programmers to pinpoint critical variables to checkpoint quickly within a few minutes. We evaluate AutoCheck on 14 representative HPC benchmarks, demonstrating that AutoCheck can efficiently identify correct critical variables to checkpoint.
\end{abstract}


\section{Introduction}

There have been many fault tolerance techniques proposed for better HPC system resilience, 
such as Algorithm-Based Fault Tolerance (ABFT)~\cite{bosilca2009algorithm}, hardware ECC and 
voltage guardbands~\cite{bacha2014using}, Dual Modular Redundancy (DMR)~\cite{vadlamani2010multicore}, and 
Checkpoint/Restart (C/R).
Among those, C/R is the most essential and widely used fault tolerance technique in HPC~\cite{laguna2014evaluating,8665769}. 
It has been deployed at almost all production HPC systems and commercial data centers by default~\cite{di2014optimization,shahzad2018craft}. 
In addition, C/R also saves state of an application in execution for time-out jobs~\cite{ba2016tool}. 
The basic idea of C/R is simple, which stores the application execution state periodically to storage and restarts from the latest stored once upon a failure.
There are plenty of C/R libraries~\cite{hargrove2006berkeley,laadan2007transparent,ni2013acr,nicolae2011blobcr}, developed at the application, compiler, and system levels, with various optimization focuses.
For example, FTI~\cite{bautista2011fti} and VeloC~\cite{nicolae2019veloc}, which are scalable application-level C/R libraries, and BLCR~\cite{hargrove2006berkeley}, which enables checkpointing to hierarchical system-level storage.

While these C/R libraries are powerful, programmers must have domain knowledge to identify which variables to checkpoint. Unfortunately, many programmers do not have such knowledge. 
On the other hand, domain scientists who have domain knowledge can still have problems identifying which variables to checkpoint without expertise in fault tolerance.
Furthermore, real-world HPC applications can be extremely complicated, composed of nested function calls, complex data structures, and intricate data dependency. 
Additionally, 
it is often a trial-and-error process, which is challenging and prone to errors, inconsistencies, and subjective interpretations,
even for professionals, to identify which variables to checkpoint.
As a consequence, there is a need for an approach and a tool that can identify which variables to checkpoint and pass over to C/R developers who are responsible for the implementation.

%


To solve this problem, we propose \textbf{AutoCheck}, an analytical model and tool that can \emph{automatically} identify \emph{critical variables} (or \textbf{\emph{the minimal set of variables necessary for checkpointing to correctly restart upon failures}}). \textit{Our solution is the first of its kind.} 
AutoCheck is a complement to existing C/R libraries. 
AutoCheck frees programmers and domain scientists from prohibitive workloads, which require understanding algorithm semantics and manually tracking \emph{complicated} data dependency across nested loops and function calls, 
via efficiently tracking dynamic data dependency and a collection of heuristics for identifying critical variables for checkpointing \emph{automatically}. 

AutoCheck's analytical model takes the dynamic application execution trace as input and has two key modules.
\emph{First}, identifying the main computation loop's input variables, for which we identify these variables defined before and used inside the main computation loop for further analysis.  
This allows us to confine the critical variables to analyze.
\emph{Second}, we track and extract the data dependency \emph{only} to those variables by \emph{selectively} iterating the dynamic instructions, 
and propose a composition of optimizations to construct a \emph{contracted} data dependency graph (DDG).
It allows us to quickly and progressively pinpoint useful information from \emph{complicated} data dependencies. 
In summary, the contributions of this paper are, 
\begin{enumerate}

\item 
A trace analysis tool that can automatically identify variables necessary for checkpointing to recover from system failures;

\item 
A selectively trace iterating model that enables us to quickly capture the data dependency between critical variables; 
\item 
The design of an analytical model and a collection of heuristics that allow identification of variables to checkpoint from an optimized data dependency abstraction;
\item 


Evaluation of AutoCheck on \emph{14} representative HPC applications, including a real-world cosmology simulation application--Hardware Accelerated Cosmology Code (HACC). 
AutoCheck successfully identifies necessary variables to checkpoint in all cases with low analytical costs: 
169.37 seconds on average with minimum 0.14 seconds and maximum 823.43 
seconds, without parallel processing, and 70.16 seconds on average with minimum 0.04 seconds and maximum 340.51 
seconds with OpenMP parallelization.
Compared to \emph{manual} trial-and-error practices, AutoCheck is \emph{easier} and \emph{faster}.


\end{enumerate}

\section{Background}
\subsection{Fault model}
C/R approaches are used to address stopping failures, due to loss of power, device worn-out, etc., and time-out jobs, which are commonly found in HPC systems and production data centers.
As reported in \cite{ostrouchov2020gpu}, the MTBF (Mean Time Between Failures) for process and node failures at flagship supercomputers has reduced to a few hours, 
while long-running, distributed HPC applications, simulations, and machine learning training can last for a few days~\cite{mergen2006virtualization,klein2017fast,zhu2011planning}. 
This causes those long-running executions to fail and stop in the middle of execution because of the loss of one or more compute nodes or processes that are not responding.


\subsection{Checkpoint/Restart model}
C/R approaches enable those long-running HPC applications to successfully restart in a fail-stop failure. 
Those long-running HPC applications typically include three parts: the preprocessing (or initialization) interface, the main computation loop, and the postprocessing (or verification) interface.
\textbf{The main computation loop refers to the part of the program that is responsible for executing the primary computational tasks. This loop is typically the core of the program and repeats a series of computational steps until a certain termination condition is met.}

To restart the main computation loop from the iteration before the failure rather than from scratch, 
we must identify what application states (or variables) are necessary  to correctly restart the application upon failures.

\emph{Past C/R approaches manually identify those critical variables for checkpointing; we provide an analytical model and tool (AutoCheck) that can automatically do it.}
Once those critical variables are identified, C/R approaches periodically store those variables to persistent storage across the iterations of the main computation loop with a certain interval. 
\textbf{A variable is defined as a memory location associated with a symbolic name, invoked in execution.} 
Upon a process or node failure, the crashed execution can restart from the latest saved checkpoint to continue from the last iteration of the main computation loop before the failure. 

\textbf{C/R insertion.}
There are two key points when inserting the C/R code once the variables to checkpoint are identified.
1) \emph{Reading checkpoints}, which is inserted right before the main computation loop to ensure all critical variables are correctly restored before the \texttt{main loop} starts.
2) \emph{Writing checkpoints}, which is added to each individual checkpointing variable within the main loop, usually at the very end of the main loop to ensure consistency. 

\subsection{Dynamic instruction execution trace}
\label{sec:llvm-trace}
\begin{figure}
  \begin{center}  \includegraphics[width=0.49\textwidth]{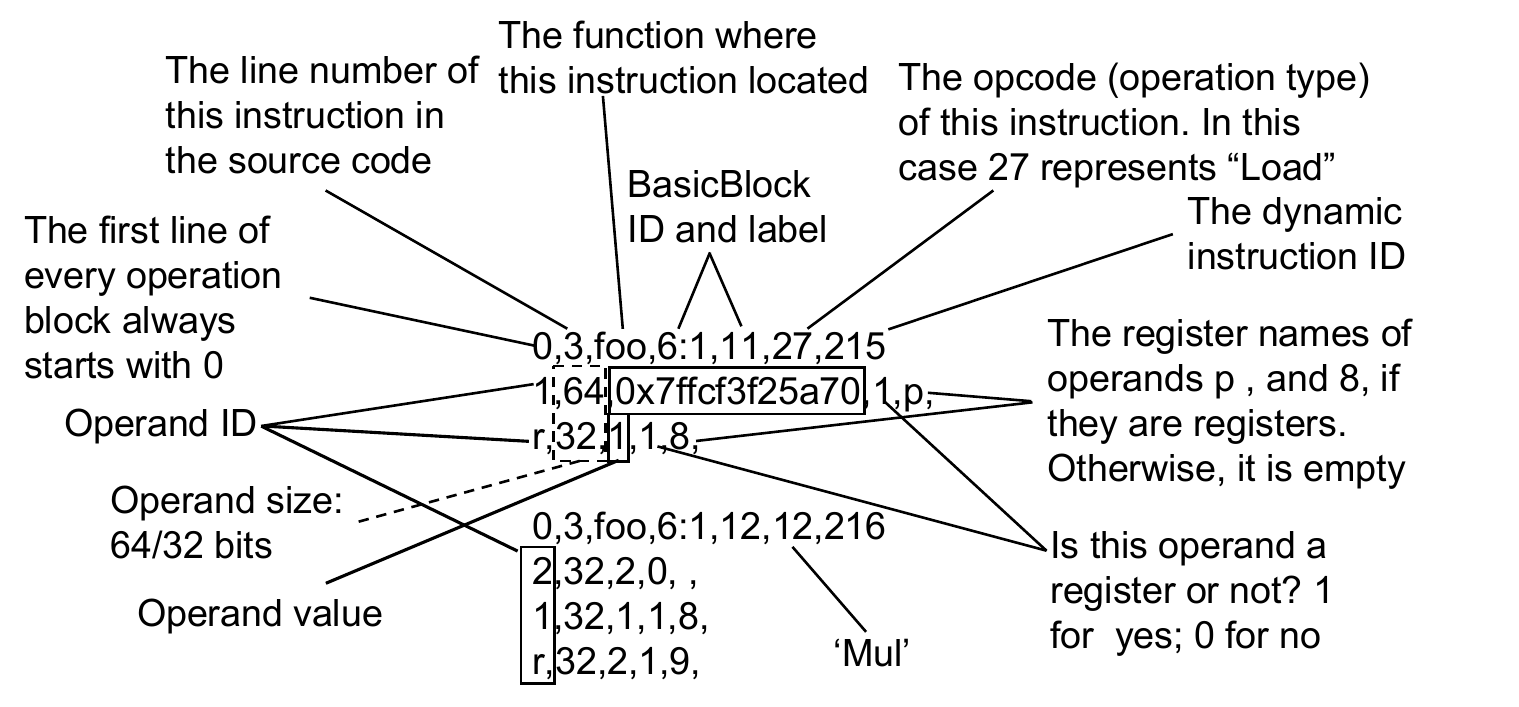}
  \caption{An example of dynamic instruction execution trace, including two instruction blocks.} 
  \label{fig:trace format}
  \end{center}
  \vspace{-2em}
\end{figure}

AutoCheck uses a dynamic instruction execution trace as input,
which is generated by an LLVM instrumentation pass, called LLVM-Tracer~\cite{shao2013isa}.
Note that we cannot rely on static analysis because of unresolved branches and data values, which are key for complete data dependency analysis. For more discussion about dynamic analysis vs. static analysis, please see Section~\ref{sec:discussion}.
LLVM-Tracer prints out a dynamic LLVM IR trace, including dynamic register values and memory addresses.
\emph{The generated instructions are LLVM IR instructions, including control-flow information and all application execution state.} 
\textit{Using LLVM-Tracer as is, we can generate the dynamic instruction execution trace for serial, OpenMP, and MPI programs while the number of OpenMP threads is one and the number of MPI ranks is one.}
The generated dynamic instruction execution trace provides detailed information for each dynamic instruction.
Figure~\ref{fig:trace format} describes a segment of the generated dynamic instruction execution trace. 
In the trace segment, there are two blocks; each block corresponds to a dynamic instruction and provides dynamic execution information of the dynamic instruction,
such as, the location of the instruction in the source code, the function where the instruction located, basic block ID and label, opcode (operation type), dynamic and operand value. 
In the first instruction, `\texttt{foo}' is the function where this instruction is located; `6:1' ($<line~number>:<column~number>$) and `11' are the basic block ID and label; `27' is the opcode that indicates the operation type which is `\texttt{Load}'; `215' is the dynamic instruction ID; `1' and `r' are the operand ID; `32' and `64' are the operand sizes; `1' indicates this operand is a register, otherwise it is `0'; `$p$' and `8' are the register names.
This instruction loads the value of variable `$p$' into temporary register `8'.
The second instruction is an `\texttt{Mul}’ instruction, which is somewhat different from the first one. This instruction multiplies the values of the operand 1 and operand 2, and stores to the temporary register `9'.



\section{Complexity of Manual Inspection}
\begin{figure*}[ht]
  \begin{center}
  \includegraphics[width=0.9\linewidth]{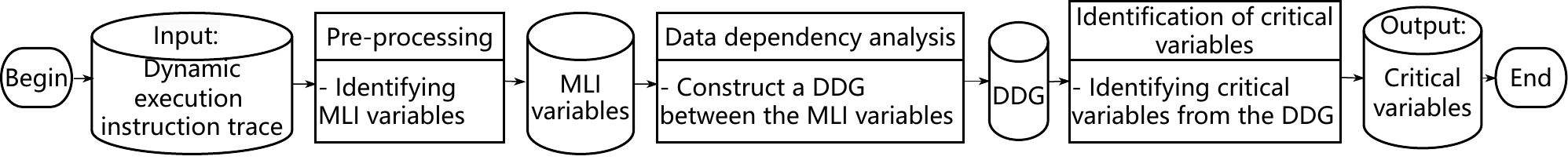}
  \caption{AutoCheck workflow diagram. }
  \label{fig:tool_workflow}
  \end{center}
  \vspace{-1em}
\end{figure*}
Manually identifying variables requires excellent patience and is prohibitively difficult. 
We summarize the complexity in three perspectives given our experience with identifying checkpointing variables for many HPC programs manually. 

\textbf{(1) Nested function calls}: Nested function calls are found in all 14 benchmarks, with depth ranging from two to as many as eight. 
For example, \textbf{AMG} has a nested function calls with depth of eight, starting with \texttt{main} and ending at \texttt{hypre\_LowerBound}, the deepest function call. 
The data dependency within \texttt{hypre\_LowerBound} played a vital role in determining checkpointing variables.

\textbf{(2) Complicated data structure}: Taking as an example \texttt{sim (SimFlatSt*)} in \textbf{CoMD}, a very complicated data structure, including multiple layers of nested pointers and function calls. 
It encompasses data structures such as \texttt{Domain}, \texttt{LinkCell}, \texttt{Atoms}, \texttt{SpeciesData}, \texttt{BasePotential}, and \texttt{HaloExchange}, which are extremely complex data structures defined across many different header files.
It turned out that
few components of \texttt{sim} involved in critical dependencies, which make it a critical variable. This is impossible to capture by eye skimming.
Similarly, \texttt{Particles} in HACC is another example. 


\textbf{(3) Convoluted data dependency}: Taking as an example \texttt{u(double****)} in \textbf{BT}, a 4D array dependent on as many as \textbf{17 other different variables} across many distinct function invocations. 
AutoCheck addresses all the data dependency on \texttt{u} and finds it involved in Write-After-Read, which requires checkpointing. 

Consequently, manually identifying variables in such complex scenarios is prohibited and prone to mistakes.
Thereby, a tool to identify critical variables is highly favorable.




\section{Design}
\label{sec:design}

AutoCheck aims to develop a tool to identify critical variables automatically. 
The AutoCheck design (depicted in Figure~\ref{fig:tool_workflow}) includes three modules: pre-processing, data dependency analysis, and identification of critical variables. 
\subsection{Identifying the \underline{M}ain-\underline{L}oop's \underline{I}nput (MLI) variables}
\label{sec:design-input-vars}

The critical variables to checkpoint are solicited from the MLI variables and the induction variables. These variables are candidates for variables to be checkpointed. An MLI variable must be defined before but used in the main computation loop.
Within the main computation loop, variables can be assigned into three categories: induction variables, MLI variables 
and local variables (defined and used within the main computation loop).
Only the induction variables that are part of the main loop are critical variables.
Local variables are re-allocated and re-initialized at every iteration, and therefore local variables need not to checkpoint. 
If a variable is defined before the main computation loop, but within a function call, we exclude it from our consideration. Because this variable is a local variable which cannot affect the global execution state.

\begin{figure}[]
  \begin{center}
  \includegraphics[width=0.45\textwidth]{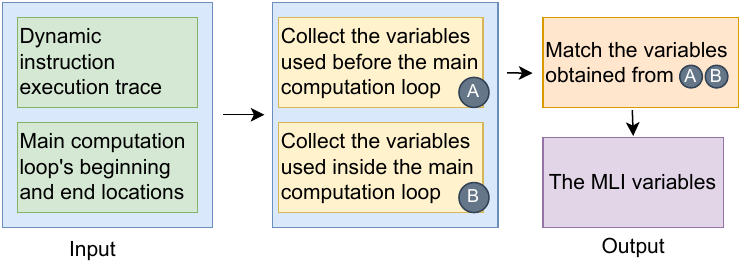 }
  \vspace{-1em}
  \caption{Pre-processing workflow.} 
  \label{fig:preprocessing_work_flow}
  \end{center}
  \vspace{-1em}
\end{figure}

This module takes the dynamic instruction execution trace and the main computation loop’s location (start and end line numbers) as input, and outputs the main-loop's input variables.
Figure~\ref{fig:preprocessing_work_flow} shows the workflow of this module.
The workflow consists of two parts: collecting the arithmetic variables used before and inside the main computation loop, and matching the collected arithmetic variables before and inside the main computation loop. 
Arithmetic variables are those variables participating in arithmetic operations.
These successfully matched variables (defined before and used inside the main computation loop) are actually the MLI variables.

\begin{figure}
  \begin{center}
    \includegraphics[width=0.43\textwidth,keepaspectratio]{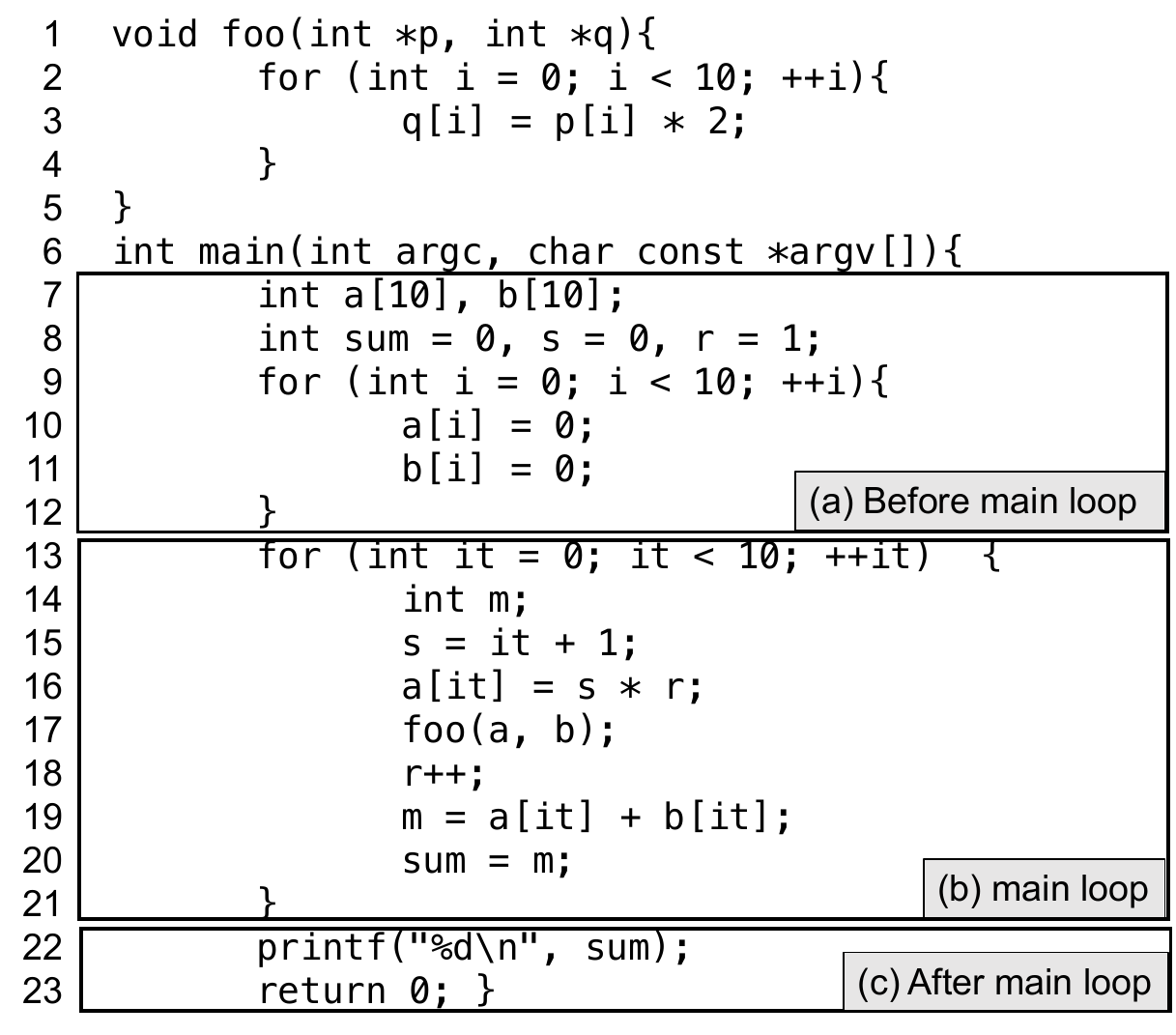}
  \vspace{-0.5em}
  \caption{Example code.}
  \label{fig:example_code}
  \end{center}
  \vspace{-1.5em}
\end{figure}

\textbf{Collect arithmetic variables:}
We collect arithmetic variables from the code region before and inside the main computation loop.
We first partition the trace into three parts: \textbf{Part A} - before the main computation loop (e.g., Region `(a)' in Figure~\ref{fig:example_code}); \textbf{Part B} - the main computation loop (e.g., Region `(b)' in Figure~\ref{fig:example_code}); and \textbf{Part C} - after the main computation loop (e.g., Region `(c)' in Figure~\ref{fig:example_code}).
We then collect arithmetic variables from \textbf{Parts A} and \textbf{B}. Note that, when \textit{POINTER ASSIGNMENT} occurs, we recursively search for the source variable in the assignment operation based on the operand of the assigned object, and replace the assigned object to collect it, which is not regarded as Write or Read.


\textbf{Match arithmetic variables:}
 Finally, we match the collected arithmetic variables from Parts A and B.
\emph{We say two variables are matched when the two variables' operand value and register name are both matched.} 
Note that, although an operand value can change during the course of its computations, the operand value does not change before participation in any  arithmetic operations as compared to its value before the main computation loop. 
Those successfully matched variables turn out to be the main computation loop's input variables, which are declared before the main computation loop and referenced within the main computation loop.
For the example code in Figure~\ref{fig:example_code}, `$a$', `$b$', `$sum$', `$s$', `$r$' are the MLI variables. 

\begin{figure*}
  \begin{center}
  \includegraphics[width=\linewidth]{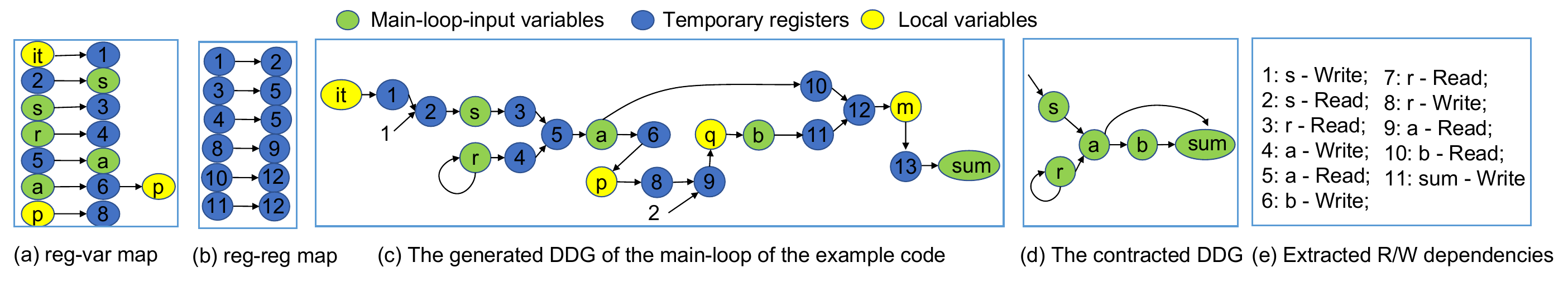}
  \caption{Data dependency analysis (R/W = Read/Write). 
  Note that reg-var map in (a) is updated on-the-fly while passing dynamic instructions. Thus, reg-var map only contains active state at a certain point.}
  \label{fig:ddg_analysis}
  \end{center}
   \vspace{-2em}
\end{figure*}

\subsection{Data dependency analysis}
\label{sec:data_dependency_analysis}
The data dependency analysis module takes the MLI variables as input, and tracks the data dependency only between MLI variables.
By doing that, we could understand the read and write dependencies on each MLI variable, and further determine which variables to checkpoint based on their read and write dependency status.  
This module eventually generates DDG between the MLI variables. In a typical application execution, when performing arithmetic operations, the variable instances are first loaded into temporary registers and these temporary registers participate in arithmetic operations, and then the result register is stored back to variables.
Therefore, we cannot generate a complete DDG between MLI
variables without understanding correlation between MLI variables and temporary registers.
Figure~\ref{fig:ddg_analysis}(c) illustrates a toy example of the generated DDG, in which `$s$', `$r$', `$a$', `$b$', `$sum$' are input variables to the main computation loop. 
We must understand how temporary registers and local variables interact with the MLI variables, 
and how to reduce a complete DDG, such as Figure~\ref{fig:ddg_analysis}(c), to 
a DDG \emph{only} including the MLI variables, such as Figure~\ref{fig:ddg_analysis}(d).
We break down this problem into two steps: 1) construct a \emph{complete DDG}, which includes the correlation between the MLI variables and other application state, including local variables and temporary registers; 
2) contract the complete DDG to a \emph{contracted DDG} that \emph{only} includes the MLI variables.


\subsubsection{\textbf{Construction of complete DDG}} 
To generate the complete DDG that includes all arithmetic variables (including the MLI variables and local variables) 
and temporary registers, we must focus on arithmetic instructions, since arithmetic variables participate in computations through arithmetic instructions.
For that, we \emph{selectively} pass the dynamic instructions, meaning that we focus the data dependency analysis on arithmetic instructions (such as `\texttt{Mul}') and the complement instructions (such as `\texttt{Load}' and `\texttt{Store}') to arithmetic instructions and transition instructions, particularly function calls. 
We propose analytical models for each of four instruction types, including 
\textit{load and store}, \textit{arithmetic}, and \textit{call}, which are the driver for computations. 

\textbf{Load and Store. }Arithmetic variables participate in arithmetic operations through temporary registers, in which arithmetic variables are first loaded to temporary variables and stored back after computation.
To associate arithmetic variables with temporary registers, 
we create a table of correlations between temporary registers and arithmetic variables.
This table maps a temporary register to the arithmetic variable which it \emph{loads} from and \emph{stores} to. 
This allows us to find the variable that participates in an arithmetic operation by looking at the table.
We describe an example of the created table in Figure~\ref{fig:ddg_analysis}(a), which ties a temporary register to a variable. 
Given a new `\texttt{Load}' instruction, for example, the first instruction block in Figure~\ref{fig:trace format},
we add the correlation between the arithmetic variable (`$p$') and the temporary register (`8') to the table.
\textbf{We call this table ``reg-var map'' in the remainder of this paper.}
Note that we update this table on-the-fly while passing dynamic instructions and following the instruction execution order.
This is because of (Static Single Assignment) SSA, in which variable instances are re-loaded again into a temporary register every time when they are re-used.
This guarantees that we associate temporary registers to the right variable.
This also resolves the ``Mutable-register'' challenge, where
a temporary register can be shared by different arithmetic variables.

\textbf{Arithmetic.}
Computations are melted down into arithmetic instructions, which are critical 
to data dependency construction. Because those instructions, such as `\texttt{Add}', tie individual temporary registers together. 
We must capture such dependencies from arithmetic instructions.
Particularly, we create a table to store the correlation between temporary registers
that links an arithmetic instruction's input and output operands that are 
temporary registers.
\textbf{We call this table ``reg-reg map'' in the remainder of this paper.}
Given an arithmetic instruction, for example, the second instruction block (`\texttt{Mul}') in Figure~\ref{fig:trace format}, 
from which we add the link between the input operand (temporary register `8') and the output operand (temporary register `9') to ``reg-reg map''.
This example is taken from the multiplication instruction within the function call \texttt{foo}() at Line 3 in 
the example code (Figure~\ref{fig:example_code}). 
Figure~\ref{fig:ddg_analysis}(b) is an example of ``reg-reg map'' for the main computation loop of the example code, in which 
input operands are linked to output operands.

\textbf{Call.}
Function calls are critical to data dependency analysis as 
they bridge the data dependency while entering and exiting the function call.
There are two forms of function calls found in the generated LLVM trace: 1) a single `\texttt{Call}' instruction; 2) a `\texttt{Call}' instruction followed by its function body.

\begin{figure}
  \begin{center}
  \includegraphics[width=0.44\textwidth]{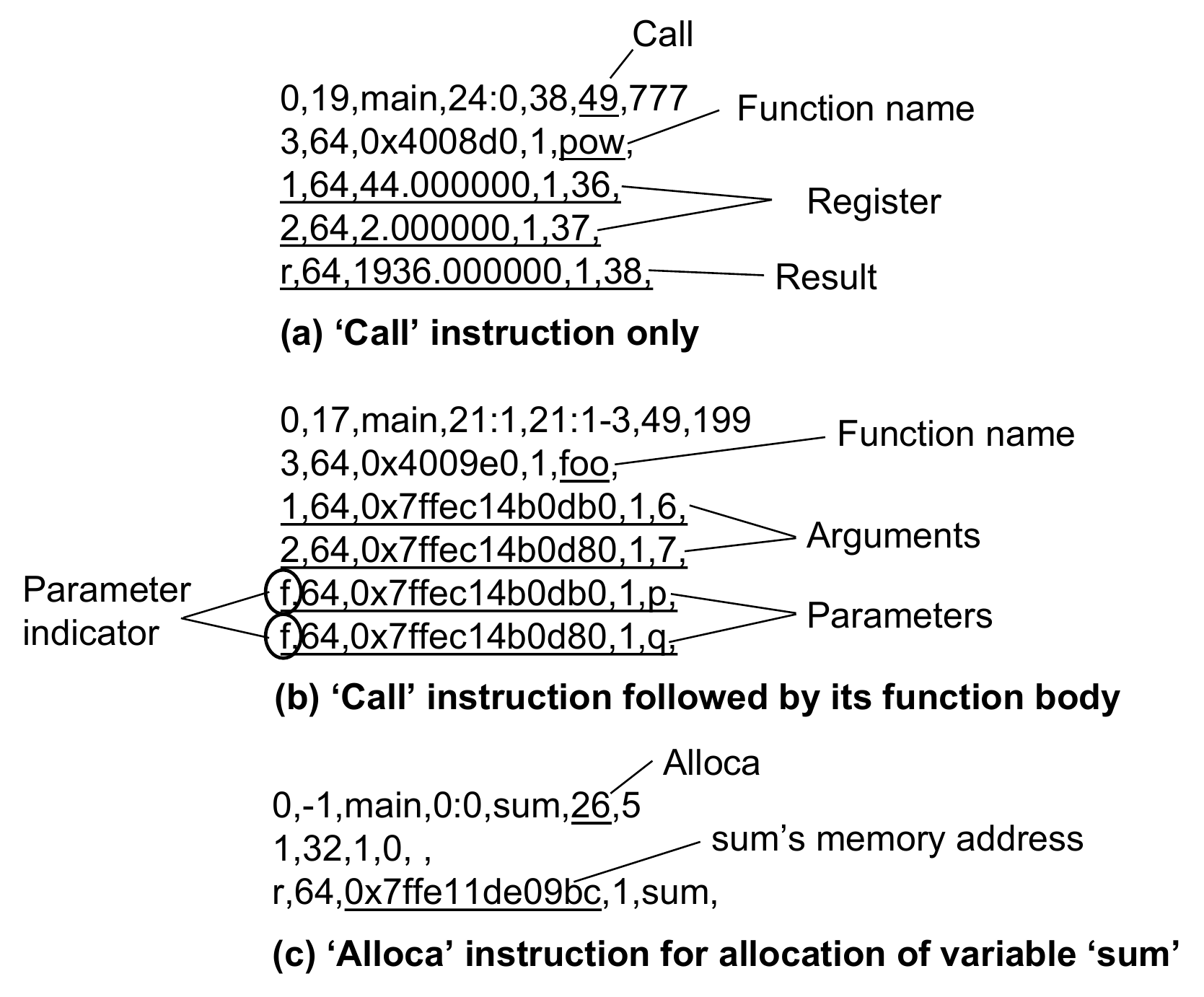}
  \caption{Critical instructions used in analysis, including two forms of `\texttt{Call}' instructions and `\texttt{Alloca}'.}
  \label{fig:llvm_instr}
  \end{center}
   \vspace{-1.2em}
\end{figure}

For 1), we treat it as arithmetic instructions. In particular, we add the correlation between input operands and output operands to the \textbf{``reg-reg map"}.
We provide an example of 1) in Figure~\ref{fig:llvm_instr}(a), which is a single `\texttt{Call}' instruction, 
for which we \emph{must} add the correlation between the registers `36' and `37', which are the input, and the register `38', which is the output, to the \textbf{``reg-reg map"}
to be involved in data dependency analysis.

The `\texttt{Call}' instruction in 2) is much different from that in 1). 
The input to the `\texttt{Call}' instruction in 2) is the arguments and the output is the parameters of the function call. 
The parameters are substituted for arguments to be used in computations of the function body. 
We provide an example of 2) in Figure~\ref{fig:llvm_instr}(b),
which is the LLVM trace of the function call \texttt{foo}() from the example code in Figure~\ref{fig:example_code}.
In Figure~\ref{fig:llvm_instr}(b), registers `6' and `7' are temporary registers loaded from 
arguments and registers `$p$' and `$q$' are parameters that are local variables. 
However, the `\texttt{Call}' instruction itself only provides correlations between the temporary registers and parameters.
In order to find the correlation between arguments and parameters, 
we must track back one instruction to find the `\texttt{Load}' instructions that 
generate temporary registers (e.g., registers `6' and `7') from arguments.
In particular, we append the correlation found in the `\texttt{Call}' instruction to the existing pairwise correlations in \textbf{``reg-var map"}, which makes triplet correlations, indicating the correlation between arguments and parameters. 
For example, for the case in Figure~\ref{fig:llvm_instr}(b), we append the correlation between `6' and `$p$' to the existing correlation in the \textbf{``reg-var map"}  (`$a$' and `6'), which constructs the correlation between `$a$' (argument) and `$p$' (parameter) through `6' (see Fig.~\ref{fig:ddg_analysis}(a)).

\textbf{Putting all together.}
The complete DDG is generated using data dependencies collected in ``reg-var map'' and ``reg-reg map''.
We update DDG with the collected  dependencies from ``reg-var map'' and ``reg-reg map''
each time at a `\texttt{Store}' instruction, 
This is because every computation
terminates by `\texttt{Store}' instructions. 
Furthermore, we summarize the instructions examined in data dependency analysis in Table~\ref{tab:instruction}, including how and where they are used.

\subsubsection{\textbf{DDG contraction}}
Our goal of data dependency analysis is to generate DDG on the MLI variables \emph{only}.
We propose an effective algorithm to contract the unnecessary vertices that are not MLI variables from the DDG. 
In particular, the algorithm takes the complete DDG from the last step as input, and replaces the parent of each
MLI variable (if it has a parent) with its grandparent (parent's parent) until all its parents become MLI variables or the DDG does not change any more. 
To the end, when the iterations break out, a MLI variable's parent remains not a MLI variable and this parent has no parents, we contract this parent vertex from DDG while retaining its dependencies. 
Eventually, all vertices in the contracted DDG become MLI variables.
We describe this in Algorithm~\ref{algo:compress_ddg}.
We describe an example in Figure~\ref{fig:ddg_analysis}, 
in which (c) is the complete DDG and (d) is the contracted DDG.
Take `$sum$' (a MLI variable) as an example, we first replace `13' with `$m$' which contracts `13' and makes `$m$' point to `$sum$' directly. 
We then contract `$m$' and make `12' point to `$sum$', after that, contract `12' and point `10' and `11' to `$sum$', and finally contract `10' and `11' and point `$a$' and `$b$' to `$sum$'. 

\begin{algorithm}[t]
  \scriptsize
  \KwIn{The complete $DDG$}
  \KwOut{The contracted $DDG$ }
 \SetKwFunction{FMain}{ContractedDDG}
    \SetKwProg{Fn}{Function}{:}{}
    \Fn{\FMain{$complete\_DDG$}}{
        \For{$n \in$ main\_loop\_input variables}{
            Get all parent vertices $NP$ of $n$\;
            \For{$np \in$ $NP$}{
                \If{$np$ is not a main\_loop\_input variable}{
                    Contract($np$);
                }
            }
            Get updated parent vertices $NP$ of $n$\;
            \For{$np \in$ $NP$}{ 
                \If{$np$ is not a main\_loop\_input variable}{
                    Contract $np$ while retaining its dependency with $n$\;
                }
            }
        }
    return $contracted\_DDG$;
    }
  \tcp{Contract each parent vertex recursively }
  \SetKwFunction{FMain}{Contract}
  \SetKwProg{Fn}{Function}{:}{}
  \Fn{\FMain{$p$}}{
    Get all parent vertices $PP$ of $p$\;
    Contract $p$, meaning to replace $p$ with its parent vertices\;
    \eIf{all vertices in $PP$ are main\_loop\_input variables \texttt{or} $PP$ is empty}{
       break;
    }{
       \For{$pp \in$ $PP$}{
            \If{$pp$ is not a main\_loop\_input variable}{
                    Contract($pp$)\;
            }
        }
       }
    }

    \caption{Contract vertices that are not MLI variables from the complete DDG.}
    \label{algo:compress_ddg}
\end{algorithm}

\begin{table}[]
\begin{center}
\caption {Instructions used in data dependency analysis.}
\label{tab:instruction}
\scriptsize
\begin{tabular}{|p{3.5cm}|p{4.5cm}|}
\hline
\textbf{Instructions} 	& \textbf{Purposes}   \\ \hline \hline
`\texttt{Load}', `\texttt{Store}',  `\texttt{BitCast}', `\texttt{GetElementPtr}'   	       & Complement instructions to arithmetic; construct the reg-var map that associates temporary registers with arithmetic variables.            \\ \hline
`\texttt{Add}', `\texttt{Fadd}', `\texttt{Sub}', `\texttt{Fsub}', `\texttt{Mul}', `\texttt{Fmul}', `\texttt{Udiv}', `\texttt{Sdiv}', `\texttt{Fdiv}'    & Arithmetic instructions: construct reg-reg map to store the correlation between temporary registers.
\\\hline
`\texttt{Alloca}'   	       & A complement instruction to arithmetic; collect local variables defined within a function call and their memory address.
\\ \hline
`\texttt{Call}'     & Construct the reg-var map that associates parameters with arguments for function calls.    \\\hline
\end{tabular}
\end{center}

\end{table}
\subsection{Identification of critical variables }


This module takes the contracted DDG as input and outputs critical variables. The contracted DDG is stored as a linked list, 
from which we pull out read and write dependencies on individual variables
by converting the contracted DDG into an execution-time-ordered sequence of read and write dependencies, according to the order of each dynamic instruction in execution. 
We give an example of the extracted read and write dependencies in execution time order in Figure~\ref{fig:ddg_analysis}(e).
With the obtained read and write dependencies, 
we can address the data dependency on individual variables to identify critical variables. 
A recent study~\cite{2024benchmarking} groups critical variables to checkpoint into \emph{four} types based on data dependency patterns.

\textbf{\underline{W}rite-\underline{A}fter-\underline{R}ead (WAR)}~\cite{tsai1996superthreaded}: This dependency pattern describes a set of data dependencies across iterations on a specific variable. 
In particular, in the first iteration, the variable is read by another variable, such as `$tmp$', and then overwritten by a new value; in the second iteration, the new value is passed to `$tmp$'. 
This means that the variable's value changes across iterations while other variables take this variable as input.
This variable needs to checkpoint; otherwise, in a failure, the variable's state can get lost, and the variable cannot be recovered without checkpoints upon a restart. 
An example is `$r$' (at Lines 16 and 18) in the example code (Figure~\ref{fig:example_code}), also illustrated in Figure~\ref{fig:ddg_analysis}(d).

\textbf{Outcome}: An outcome is the main-loop's output, which is used in followup computations as input after the main computation loop. An example is `$sum$' (at Line 20) from the example code illustrated in Figure~\ref{fig:ddg_analysis}(d). 

\begin{figure}
    \centering \includegraphics[width=0.60\linewidth]{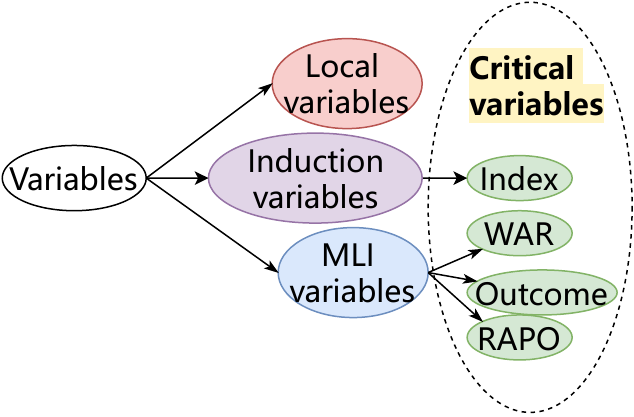}
    \caption{Identifying critical variables for checkpointing. }
    \label{fig:variablestype}
\end{figure}

\textbf{\underline{R}ead-\underline{A}fter-\underline{P}artially-\underline{O}verwritten (RAPO)}: When a variable is an array, a write to it can result in partial overwriting, where only some of its elements are updated before they are read by others. In the event of a failure during restart, the elements that were not involved in the overwriting cannot be recovered. To maintain the state of the array, it is necessary to checkpoint it.

An example is `$a$' (at Line 16) in the example code illustrated in Figure~\ref{fig:ddg_analysis}(d).

\textbf{Index}. 
We also do checkpoint to the induction variables of the main computation loop (only the outermost loop of the main computation loop). And we use llvm-pass-loop API for identification.
This is a separate case.
The reason is that in a restart, we can direct the execution to the iteration where it dropped out because of the failure.
For example, as shown in Figure~\ref{fig:example_code}, Line 13 is the `for' conditional statement of the main computation loop, in which `$it$' is the induction variable to checkpoint. 

The critical variables for checkpointing are summarized in Figure~\ref{fig:variablestype}. Consequently, for the example code in Figure~\ref{fig:example_code}, whose DDG is in Figure~\ref{fig:ddg_analysis}(d), based on the proposed principles, we should checkpoint variables `$r$', `$a$', `$sum$' and `$it$'.

\subsection{Case Study: CG}

\begin{algorithm}[t]
  \scriptsize
  \caption{CG pseudocode}
  \label{alg: pseudo code}
  \KwIn{$x, z, p, q, r, A$}
  \KwOut{$rnorm$}
  \tcp{Temporary variables: $\alpha$, $\beta$, $\rho$, $\rho_{0}$}
  \SetKwFunction{FCG}{conj\_grad}
  \SetKwProg{Fn}{Function}{:}{}
  \Fn{\FCG{$x, z, p, q, r, A$}}{
    $z = 0$\\
    $r = x$\\
    $\rho = r^Tr$\\
    $p = r$\\
    \For{$each$ $iter \in$ $cgitmax$}{
      $q = Ap$\\
      $\alpha$ = $\rho$ / $(p^Tq)$\\
      $z = z + \alpha p$\\
      $\rho_{0} = \rho$\\
      $r = r - \alpha q$\\
      $\rho = r^Tr$\\
      $\beta = \rho / \rho_{0}$\\
      $p = r + \beta p$\\
    }
    \Return $rnorm = ||x - Az||$
  }
  \KwIn{$x, z, p, q, r, A$}
  \KwOut{$rnorm, zeta$}
  \tcp{Constant variables: $SHIFT$}
  \SetKwFunction{Mainloop}{main}
  \SetKwProg{Fn}{Function}{:}{}
  \Fn{\Mainloop{$x, z, p, q, r, A$}}{
    \tcp{The main loop}
    \For{$each$ $iter \in$ $NITER$}{
      \textit{$rnorm = conj\_grad(x, z, p, q, r, A)$}\\
      $x = z / \sqrt{z^Tz}$\\
      $zeta = SHIFT + 1/x^Tz$
    }
  }
  \tcp{R/W dependencies of the main loop}
   Line 2: z-Write;\\
   Line 3: x-Read, r-Write;\\
   Line 5: r-Read, p-Write;\\
   Line 7: A-Read, p-Read, q-Write;\\
   Line 9: q-Read, r-Read, p-Read, z-Read, z-Write;\\
   Line 11: q-Read, r-Read, p-Read, r-Write;\\
   Line 14: r-Read, p-Read, p-Write;\\
   Line 19: z-Read, x-Write;
\end{algorithm}

We show the pseudocode of CG in Algorithm \ref{alg: pseudo code}, in which the $conj\_grad$ function is declared and invoked in the \texttt{for-loop} within the $main$ function. 
Note that the input to the $main$ function are global variables, initialized at the beginning of the $main$ function before the \texttt{for-loop}. 
We use the \texttt{for-loop} in the $main$ function as the main loop for analysis. 
The read and write dependencies corresponding to specific lines are provided in Algorithm \ref{alg: pseudo code} (after Line 20), generated by data dependency analysis. 
We observe a write-after-read on $x$ (read at Line 3 and write at Line 19), which thus must be checkpointed.  
For the remaining input variables, including $z$, $p$, $q$, $r$, and $A$, we did not find a dependency necessary for checkpointing.
Additionally, the indexation $iter$ must be checkpointed.




\section{Implementation}
\label{sec:impl}

\subsection{Trace analysis optimization}
\label{sec:impl-tracegen}

AutoCheck uses LLVM-Tracer to generate the dynamic instruction execution trace. 
However, the generated trace for HPC application can be very large for computation-intensive code.
To advance the trace processing efficiency,
we enable parallel processing to the input trace file using OpenMP.
In particular, we first use the master thread to partition the input file stream into sub-file-streams while not breaking individual instruction blocks (see Figure~\ref{fig:trace format}) into two sub-file-streams.
We then use worker threads to process and read many sub-file-streams concurrently
(48 OpenMP threads used in our evaluation with 16 times average speedup).
We allow users to determine whether to enable acceleration or not.

\subsection{Identifying MLI variables}
As described in Section~\ref{sec:design-input-vars}, we
identify the MLI variables by matching arithmetic variables
collected from code regions before and inside the main computation loop.
However, there are several challenges to resolve in implementation.

\textbf{Challenge 1: Mismatch of local variables.}
While identifying the MLI variables, we match the collected arithmetic variables from the code region before the main computation loop and the code region inside the main computation loop.
This can lead to a problem when function calls assigning the same name to local variables are invoked both before and inside the main computation loop.
In that specification, local variables with the same name would appear in both code regions before and inside the main computation loop, and those local variables will be matched and identified as MLI variables \emph{unexpectedly}. 
It is obviously wrong to identify local variables defined within function calls to be the MLI variables.

Our \textbf{solution} is that, when we collect arithmetic variables from the main computation loop, we bypass the code interval 
of function calls such that we exclude arithmetic variables within those function calls for the code region inside the main computation loop. 
Thus, local variables defined within those function calls won't be matched after that. 
Unfortunately, bypassing those function calls can lead to bypassing global variables created outside of functions while used in those function calls.
In our cases, we found that FT defines four global variables, $sums$, $twiddle$, $xnt$, and $y$, but uses them in function calls within the main computation loop (Lines 44 to 49 in \texttt{appft.c} from the FT source code).
As a work-around solution in this scenario, we manually initialize those global variables right before the main computation loop, to be able to include those variables into the MLI variables. 
While global variables are handy, minimizing them promotes modularity, understandability, and maintainability, benefiting code quality~\cite{lohse1984experimental,bourque2002fundamental}. 

\subsection{Data dependency analysis}

We encounter a few challenges for data dependency analysis in implementation. 
We describe those challenges and our solutions as follows.

\textbf{Challenge 2: Local variables share the same name as MLI variables.}
We have seen cases in which local variables defined in function calls can share the same name as MLI variables. 
In order to generate the correct contracted DDG, which should only include MLI variables, 
we must address this challenge to be able to determine whether a variable is a MLI variable or a deceiver that is actually a local variable.

Our \textbf{solution} to this challenge is based on an observation: all local variables defined within a function call are first allocated with `\texttt{Alloca}' instructions that allocate memory on the stack.

We provide an example of `\texttt{Alloca}' in Figure~\ref{fig:llvm_instr}(c), which allocates memory for the variable `$sum$'.
From those `\texttt{Alloca}' instructions, we can identify every local variable used in the function call and their memory addresses.
With all variables having its memory address, we could easily determine if a variable is a local or MLI variable by its memory address. 
In particular, if we can find a match between the variable's memory address and any MLI variable's memory address, the variable is a MLI variable; otherwise, the variable is a local variable.

Note that we could capture a MLI variable's memory address from its `\texttt{Load}' and `\texttt{Store}' instructions.
Also note that we cannot rely on `\texttt{Alloca}' to identify MLI variables because from `\texttt{Alloca}' instructions, 
we can identify all variables defined within the main computation loop, but cannot distinguish variables defined before versus inside the main computation loop.

\begin{table*}[h!]
\begin{center}
\caption {Benchmarks and applications for the study. OMP means OpenMP; MCLR means ``\underline{M}ain \underline{C}omputation \underline{L}oop \underline{R}ange`` ;LOC means ``\underline{L}ines \underline{O}f \underline{C}ode''. Our validations on those benchmarks are all successful.}
\label{tab:benchmark}
\scriptsize
\begin{tabular}{|p{1.1cm}|p{3.9cm}|p{0.5cm}|p{0.5cm}|p{1.2cm}|p{5.8cm}|p{2.1cm}|}
\hline
\textbf{Name} 	& \textbf{Benchmark description} 	& \textbf{LOC}  & \textbf{Trace size} & \textbf{Trace generation time (s)}	& \textbf{Critical variables    (Dependency type)}  	&\textbf{MCLR (File name)}             \\ \hline \hline
Himeno     MPI       &Measuring a CPU performance of floating point operation by a Poisson equation solver (input size $8 \times 8 \times 8$)   &243  &52M  &2.91   &$p$ (WAR), $n$ (Index)  &186-217 (himenobmt.c)          \\  \hline
HPCCG      OMP+MPI       &Conjugate Gradient benchmark code for a 3D chimney domain (input size $2 \times 2 \times 2$)  &3415   &2.6M  &0.11  &$t1$ (WAR), $t2$ (WAR), $t3$ (WAR), $r$ (WAR), $x$ (WAR), $p$ (WAR), $rtrans$ (WAR), $k$ (Index) &118-146 (HPCCG.cpp)   \\  \hline
CG (NPB) OMP           & Conjugate Gradient with irregular memory access (input size $10 \times 8 \times 2$)  &1353  &54M   &2.28     & $x$ (WAR), $it$ (Index)   &296-330 (cg.c)       \\  \hline
MG (NPB)  OMP	       & Multi-Grid on a sequence of meshes (input size $3 \times 3 \times 3$)          &1677  &98M   &4.23     & $u$ (WAR), $r$ (WAR), $it$ (Index)   &259-269 (mg.c) 	\\ \hline
FT (NPB) OMP         & Discrete 3D Fast Fourier Transform (input size $8 \times 8 \times 8$)       &1309  &213M     &9.11      & $y$ (WAR), $sum$ (Outcome), $kt$ (Index)     &101-111 (appft.c)   \\ \hline
SP (NPB) OMP          & Scalar Penta-diagonal solver (input size $3 \times 3$)         	&3570  &42M	&2.21  & $u$ (WAR), $step$ (Index)    &184-190 (sp.c)  \\ \hline
EP (NPB) OMP             & Embarrassingly Parallel (input size 3)         	&625  &1.3G	 &59.75   & $sy$ (WAR), $q$ (WAR), $sx$ (WAR), $k$ (Index)  &168-213 (ep.c) \\ \hline
IS (NPB)  OMP           & Integer Sort, random memory access (input class 4096)         &981   &367M	&16.32	& $passed\_verification$ (WAR), $key\_array$ (RAPO), $bucket\_ptrs$ (RAPO),  $iteration$ (Index)  &787-791 (is.c)    \\ \hline
BT (NPB) OMP       & Block Tri-diagonal solver (input size $3 \times 3 \times 0.0008$)     &4216  &58M    &2.76		& $u$ (WAR), $step$ (index)	  &180-186 (bt.c) \\ \hline
LU (NPB) OMP     & Lower-Upper Gauss-Seidel solver (input size $5 \times 5 \times 5$)        &4227  &1.6G   &81.96 	& $u$ (WAR), $rho\_i$ (WAR), $qs$ (WAR), $rsd$ (WAR), $istep$ (Index)	 &115-267 (ssor.c) \\ \hline
CoMD (ECP) OMP+MPI      & A proxy application in molecular dynamics (MD) often used for particle motion simulations (input size -$x$ 4 -$y$ 4 -$z$ 4)     &5637  &3.4G    &59.37	&$sim$ (WAR), $perfTimer$ (WAR), $iStep$ (Index)	&113-126 (CoMD.c)  \\ \hline
miniAMR (ECP) OMP+MPI        & A large-scale 3D stencil calculation by Adaptive Mesh Refinement (input size --$nx$ 2 --$ny$ 2 --$nz$ 2 --$max\_blocks$ 2)        &11531  &2.3G   &39.32  &29 $timers$ (WAR), $counter\_bc$ (WAR), $total\_fp\_adds$ (WAR),  $total\_blocks$ (WAR), $total\_fp\_divs$ (WAR), $total\_red$ (WAR),  $nrs$ (WAR), $nrrs$ (WAR),$num\_moved\_coarsen$ (WAR),  $num\_moved\_rs$ (WAR), $num\_comm\_uniq$ (WAR),  $num\_comm\_tot$ (WAR), $num\_comm\_z$ (WAR),  $num\_comm\_y$ (WAR), $tmax$ (WAR),$tmin$ (WAR),$global\_active$ (WAR),  $num\_comm\_x$ (WAR), $blocks$ (WAR), $done$ (Index),  $ts$ (Index)	&67-160 (driver.c) \\ \hline
AMG (ECP) OMP+MPI  & Algebraic Multi-Grid linear system solver for unstructured mesh physics packages (input size $-problem$ 2 $-n$ $ 5 \times 5 \times 5$)          &75000  &6.8G   &117.39     & $diagonal$ (WAR), $cum\_num\_its$ (WAR), $cum\_nnz\_AP$ (WAR), $hypre\_\_global\_error$ (WAR), $final\_res\_norm$ (Outcome), $j$ (Index)   	&462-553 (amg.c) \\ \hline

HACC OMP+MPI  & The Hardware Accelerated Cosmology Code framework(input size $-N$ 1 $-t$ $ 1 \times 1 \times 1$)       &32254  &12.7G   &201.13     & particles (WAR),step (Index)   	&318-523 (driver\_hires-local.cxx) \\ \hline

\end{tabular}
\end{center}
\vspace{-20pt}
\end{table*}

\begin{table}[]
\centering
\scriptsize
\caption{Efficiency study on 14 benchmarks. Optimization means parallel execution with 48 OpenMP threads.}
\begin{tabular}{|p{0.85cm}|p{1.5cm}|p{1.4cm}|p{1cm}|p{1.8cm}|}
\hline
\textbf{Name}  	 & \textbf{Pre-processing (With optimization) (s)}	& \textbf{Dependency Analysis (s)} 	& \textbf{Identify Variables (s)}	 & \textbf{Total Time (With optimization) (s)}   \\ \hline \hline
Himeno     & 5.44  (0.32)   &2.53      &5e-3     & 7.98 (2.86) \\\hline 
HPCCG      & 0.11  (0.01)   &0.03      &2e-3      &0.14 (0.04) \\\hline 
CG         & 3.17 (0.2)   &1.18      &0.02      &4.37 (1.4)       \\\hline
MG        & 5.35 (0.33)   &1.12     &0.01     &6.48 (1.46)       \\\hline
FT        & 15.93 (0.95)  &7.01      &0.06     &23 (8.02)       \\\hline 
SP        & 2.19 (0.13)  &0.55      &0.02      &2.76 (0.7)       \\\hline
EP        & 113.18 (6.94)   &90.18   &2.38     &205.74 (99.5)  \\\hline
IS         & 33.06 (2.04)  &11.57      &0.05      &44.68 (13.66)   \\\hline
BT         & 3.57 (0.22)  &1.54     &0.04      &5.15 (1.8)       \\\hline
LU         & 185.07 (12.57)  &90.66      &0.39      & 276.12 (103.6)      \\\hline
CoMD     & 129.47  (7.94)   &2e-3      &5e-4     & 129.47 (7.94) \\\hline 
miniAMR      & 142.97  (8.66)   &105.23      &0.12      &248.32 (114.01) \\\hline 
AMG      & 327.64  (20.74)   &261.43      &4.59      &593.66 (286.76) \\\hline
HACC   & 514.45 (31.53)         &301.06  &7.92   &823.43(340.51)       	\\ \hline

\end{tabular}
\label{tab:time}
\vspace{-15pt}
\end{table}

\begin{table}[]
\centering
\scriptsize
\caption{Storage cost for checkpointing.}
\begin{tabular}{|p{0.85cm}|p{2.3cm}|p{1.8cm}|p{1.9cm}|}
\hline
\textbf{Name}    & \textbf{Input size}	 & \textbf{BLCR~\cite{hargrove2006berkeley} (MBs)}	& \textbf{AutoCheck (MBs)}   \\ \hline \hline
Himeno   &$129 \times 65 \times 65$    & 32550.76   &2.53       \\\hline
HPCCG    &$64 \times 64 \times 64$  & 452202.50   &610.9       \\\hline 
CG       & S  & 16569.47   &0.16            \\\hline
MG      & S  & 3220.39   &2.84           \\\hline
FT      & S  & 53616.26  &24.6           \\\hline 
SP     & S   & 20068.88  &7.81           \\\hline
EP     & S   & 50061.67   &0.03     \\\hline
IS      & S   & 952.85  &2.53         \\\hline
BT      & S   & 34042.18  &4.69            \\\hline
LU     & S    & 17263.79  &9.33            \\\hline
CoMD   & -$x$ 8 -$y$ 8 -$z$ 8  & 375798.50   &241.71       \\\hline 
miniAMR & --$nx$ 8 --$ny$ 8 --$nz$ 8 --$max\_blocks$ 8     & 30310.90   &0.09       \\\hline 
AMG & -problem 2 -n 40 40 40     & 647577.68   &0.01      \\\hline
HACC           &-$N$ 1 -$t$ $1 \times 1  \times 1$  &217533.14   &134.93       	\\ \hline
\end{tabular}
\label{tab:space}
\vspace{-15pt}
\end{table}

\section{Evaluation}

\subsection{Experimental setup}

\textbf{Platforms}: All experiments are run on a Linux server consisting of 32 GB of DDR4 memory, 500 GB SSD, and two Intel Xeon E5-2678 V3 CPUs, each with two sockets, including 12 physical cores at 2.50 GHz frequency on each socket, and two threads per core.
All experiments (excluding LLVM trace generation) are compiled with gcc-7.5 and OpenMP-4.5 with `-O3' optimization. 

\textbf{LLVM trace generation}:
We use LLVM-Tracer 1.2 for LLVM trace generation with LLVM/Clang-3.4.2.
To generate the trace of the entire program including all function calls, following LLVM-Tracer's instructions, we set the environment variable \texttt{WORKLOAD} to
`main'
and enable the flag `-trace-all-call' in compilation.
Table~\ref{tab:benchmark} also shows size of the generated traces, times for trace generation, and the input size used for LLVM trace generation.

\textbf{Benchmarks}: 
We evaluate AutoCheck on 14 HPC applications, including HPCCG~\cite{richards2020quantitative}, Himeno~\cite{himeno}, and all of the NAS parallel benchmarks (NPB)~\cite{bailey1993parallel}, and three large HPC proxy applications, CoMD, miniAMR, and AMG, from the Exascale Computing Project (ECP) proxy application suite~\cite{sultana2021understanding}, and a real-world cosmology simulation application, HACC~\cite{habib2013hacc}.
Table~\ref{tab:benchmark} describes the 14 benchmarks, including benchmark description and lines of code (LOC) and which file the main computation loop is located including line numbers (MCLR). 
All those benchmarks are either OpenMP or MPI. We deliberately chose 14 programs to encompass a wide spectrum of domains, rather than relying solely on a single benchmark suite~\cite{guo2020match}.

\textbf{HACC}: The \underline{H}ardware \underline{A}ccelerated \underline{C}osmology \underline{C}ode (HACC) framework employs N-body methods for simulating an expanding universe. Its primary scientific objective is to model the universe's evolution from its early stages, aiming to advance our comprehension of dark energy and dark matter, which constitute 95\% of the universe.

 


\textbf{C/R library}: 
We use an actively maintained, easy-to-use C/R library, Fault Tolerant Interface (FTI)~\cite{bautista2011fti}, to verify the correctness of the variables identified by AutoCheck. 
FTI supports multiple levels of reliability and performance through local storage and data replication.
We use the most basic FTI checkpointing mode L1 for checkpointing. 
L1 mode writes checkpoints locally to a compute node.


\textbf{Code}: All the code
are included in the artifact and freely available at a git repository~\footnote{https://github.com/zRollman/Autocheck.git}.

\subsection{Validation and characterization of variables to checkpoint}
\label{sec:validation}
Table~\ref{tab:benchmark} presents the identified variables by AutoCheck and their dependency type for checkpointing. 
To validate the effectiveness of those detected variables, 
we manually add C/R code to those variables in their source code using FTI. 
To simulate a fail-stop failure, we raise an abnormal termination signal in the main computation loop by `raise(\texttt{SIGTERM})'.
It turns out that all the 14 benchmarks restart successfully, 
meaning that the output of the restart execution matches the output of a failure free execution, 
with the AutoCheck-detected variables checkpointed, after an abnormal termination. 
This justifies the \emph{sufficiency} of the detected variables.

Also, to check if AutoCheck provides unnecessary (false-positive) variables for checkpointing, we disable C/R to those variables one at a time,
and see if the execution can still restart successfully and generate correct output in this case. 
To the end, we didn't find unnecessary (false-positive) variables in the detected variables.
In addition to the 14 benchmarks, 
we haven't seen a case where AutoCheck fails.
We are confident that it will work on many more applications.

Furthermore, we aim to understand the data dependency on those variables.
We report the dependency type (one of the four dependency types including ``index") to each of those variables in the last second column of Table~\ref{tab:benchmark}. 
There are in total 102 
variables to checkpoint. 
\textit{Within the 95, 76 
are Write-After-Read; 2 is Outcome; 2 are Read-After-Partially-Overwritten; and 15 are Index.} 
This indicates that Write-After-Read is a critical dependency pattern for checkpointing.

\subsection{Efficiency study}
\textbf{Analytical cost.} Table~\ref{tab:time} shows the cost of our analytical model, including the overall analysis time and breakdowns, with/without OpenMP parallelization for pre-processing optimization (see Section~\ref{sec:impl-tracegen}: Trace Analysis Optimization). 
Each test is averaged over five runs. 
The total analysis time is up to 39.53 minutes(16.38 minutes with optimization). 
This is very efficient, in contrast to manual examination of variables for checkpointing from a parallel application within thousands lines of code. 
The most time-consuming part is pre-processing, which reads individual instructions from the LLVM trace file.
Given an LLVM trace file of size ranging from 2.6M to 12.7G, the pre-processing can take up to 8.6 minutes before optimization and 31.6 seconds with OpenMP parallelization. 
For large benchmarks, CoMD, miniAMR,  AMG and HACC, the total analysis time for them is 2.2 minutes (7.9 seconds with optimization), 4.1 minutes (1.9 minutes with optimization) , 9.9 minutes (4.8 minutes with optimization) and 13.8 minutes (5.7 minutes with optimization), respectively, although they have a much larger codebase. 
We observe that CoMD has a larger trace, but the analysis cost is much lower than miniAMR.
We look into the source code of CoMD and find out that there are more than 95\% instructions for initialization and logging and only less than 5\% instructions for the main computation loop.
AMG has a large iterative solver which calls a few compute-intensive interfaces, including the preconditioner, relaxation, restriction, and interpolation phases. 
HACC has an even larger trace than AMG and takes longer to process. 
We observed that the time complexity of our analysis tool is linear to the size of the trace (number of instructions) 
and we can easily reduce the analysis time with parallelization and more resources.
It will be our \textbf{future work} to implement AutoCheck as an LLVM instrumentation tool to cut the trace processing cost.

\textbf{Storage cost for checkpointing.}
We compute and compare the storage cost (for holding checkpoints) of AutoCheck with a commonly used system-level checkpointing C/R libraries, 
Berkeley Lab Checkpoint/Restart (BLCR)~\cite{hargrove2006berkeley},  
on all 14 benchmarks, with the same input. 
We provide calculated storage cost in both cases in Table~\ref{tab:space}.
We can observe that the storage cost is significantly lower (up to seven orders of magnitude) in our case as compared with BLCR, 
which demonstrates AutoCheck's storage efficiency.
\section{Discussion}
\label{sec:discussion}



\textbf{Use of AutoCheck.}
To use AutoCheck, one needs to give AutoCheck correct inputs, including 1) the dynamic execution trace of the target program generated by LLVM-Tracer with a small input (for better efficiency), 2) the main computation loop's start and end line numbers, and 3) the name and location (line number) of the function where the main computation loop is.
\textbf{Note that AutoCheck is generally applicable to any block of continuously executed code as soon as we know its start and end locations, beyond the main computation loop.}
The output of AutoCheck includes 1) name of the variables to checkpoint and 2) the location (line numbers) of declaration of those variables. 

\textbf{Select main loop.}
The main loops in the 14 benchmarks are found manually, and they are the most computationally intensive and longest running loops. AutoCheck works with any loop, but each loop generates different checkpoint variables. In cases where it is difficult to identify the main loop, it is possible that there are multiple loops. In this case, AutoCheck can process one loop at a time.

\textbf{With different inputs.} 
Normally variables to checkpoint do not change for different inputs. 
This is also common sense in C/R and how existing C/R libraries work. 
Nobody changes variables to checkpoint every time for different inputs. 
In fact, once people add the C/R code to programs, they don't change it anymore.
We also check if this holds for our benchmarks. In particular, we tested the AutoCheck-detected variables for checkpointing on larger input problem sizes, different from the small input we used for AutoCheck analysis, for all the 14 benchmarks. 
It turns out that, in all cases, the restart execution, taking different, larger inputs, runs successfully, with the same detected variables for checkpointing. 
Despite that, we still encourage users to run AutoCheck every time when the input is different.

\textbf{MPI programs.}
AutoCheck works for MPI programs with one rank as it works for serial programs. 
We assume that all MPI processes adhere to Bulk Synchronous Parallel (BSP) execution model.
Basically, checkpointing in parallel applications is a synchronous one where MPI processes start checkpointing after global MPI collective communications, such as \texttt{MPI\_Barrier}, to eliminate inter-process dependency and avoid \emph{Domino effect}~\cite{Baldoni1995OnMC}.
Thus, in synchronous checkpointing, the process is localized, negating the requirement for inter-process dependency analysis.
On top of that, AutoCheck works for asynchronous checkpointing. Essentially, communication is an operation copying one buffer on a node to another buffer (i.e., fully or partially overwritten) on a different node. Thus, when we analyze dependency for how this buffer is used in each node, our dependency analysis has considered inter-process dependency already by itself. 
To summarize, 1) all the checkpointing variable detection is local work, thus we don’t need to consider other processes; 
2) our approach also considers the communication buffer and how this buffer depends on which variable, so our analysis already considers inter-process dependency.

\textbf{OpenMP programs.} OpenMP works in our case. Not every thread writes checkpoints. Only the master thread writes checkpoints on behalf of other threads. Because outside the parallel region, it should be single threading. Parallel regions should be within the main computation loop. All the reading and restarting from checkpoints are before the main computation loop. This is totally single threading. If you write a checkpoint within a parallel region, the master thread can write the checkpoint locally.

\textbf{Parallel and Serial.}
AutoCheck relies on the data dependency among the main loop input (MLI) variables to identify checkpointing variables, meaning it only seeks checkpointing variables from MLI variables. Even though new local variables may be introduced within the main loop by models like Bulk Synchronous Parallel (BSP) and non-BSP, along with complex communication patterns, the algorithmic meaning should remain consistent with the serial code to generate correct results. Consequently,the data dependencies between MLI variables should remain constant in both the serial and parallel cases.

\textbf{Limitations.} Currently AutoCheck only works for C/C++ programs because 
AutoCheck is built on LLVM/Clang techniques, which only support code generation for the C language family.
As a result, we only evaluate AutoCheck on 14 benchmarks in C/C++ with OpenMP and MPI. 
We keep it for future work to extend the AutoCheck approach to more programming languages and libraries.  

\section{Related work}

\textbf{Checkpoint/Restart.} 
There have been a variety of C/R libraries~\cite{bautista2011fti, hargrove2006berkeley,laadan2007transparent,ni2013acr,nicolae2011blobcr,nicolae2019veloc,ansel2009dmtcp,guo2020match,georgakoudis2020reinit} developed with different focuses. 
Bautista-Gome et al.~\cite{bautista2011fti} developed Fault Tolerant Interface (FTI), which is a scalble, multi-level checkpointing library.
Nicolae et al.~\cite{nicolae2019veloc} provided an asynchronous checkpointing solution. 
However, all those C/R libraries request programmers with little domain knowledge to identify variables to checkpoint manually.
To solve this problem, we develop a tool, AutoCheck, that leverages data dependency analysis and a series of heuristics to automatically identify variables for checkpointing.

\textbf{Data-dependency-based resilience analytic.} 
Numerous existing research~\cite{li2016understanding,sastry2014ganges,li2008understanding, li2018modeling,li2022visual,guo2019moard,guo2021paris,guo2018fliptracker,guo2023understanding,fu2023high} leverages data dependency analysis to do resilience analytics and to identify code vulnerabilities.
Guo et al.~\cite{guo2019moard} developed an analytical tool for measuring application vulnerability on individual variables by counting error masking events from data-dependency analysis.
Menon et al.~\cite{menon2018discvar} leveraged Automatic Differentiation (AD) for data correlation analysis to identify resilient variables to transient faults. 
In contrast, our work aims to free programmers from manually tracking complicated data dependency across loops and functions to identify variables for checkpointing. 
Our work extends the capability of C/R libraries.

\textbf{Automated checkpointing.}
There have been a few notable research efforts focusing on automated checkpointing techniques. For example, compiler-assisted automated checkpointing instruments the application with C/R code automatically~\cite{zhao2012compiler,bari2020checkpointing,liu2016compiler,vogt2015lightweight,bronevetsky2008compiler,plank1995compiler}. 
The other is automated application-level checkpointing~\cite{bronevetsky2003automated,ba2016tool,shahzad2018craft,bronevetsky2004c,bronevetsky2004application,takizawa2017application}, which enables checkpointing by automatically preserving specific application state (or processes). 
In addition, there are automated system-level checkpointing methods, which save the state of the entire system to checkpoints automatically~\cite{agarwal2004adaptive,plank1997overview,cao2016system,di2014optimization,kulkarni2012design}.
However, to the best of our knowledge, there is not an existing approach that can identify variables to checkpoint for C/R automatically.
AutoCheck is the first work that can identify critical variables necessary for checkpointing.
However, there is no tool or method to automatically recognize C/R checkpoint variables.



\section{Conclusion}
This paper introduces AutoCheck that can identify variables for checkpointing automatically.
We demonstrated its effectiveness and efficiency with 14 HPC applications.
In future work, 
we aim to enable C/R code to the detected variables for checkpointing automatically within AutoCheck.
In particular, we attempt to incorporate AutoCheck into LLVM to be an independent LLVM instrumentation tool to eliminate the performance bottleneck because of trace file processing.
This also extends AutoCheck's capability to deal with various programming languages and models beyond OpenMP and MPI.

\bibliographystyle{IEEEtran}
\bibliography{main}

\end{document}